\documentclass[conference, 10pt, a4paper]{IEEEtran}
\IEEEoverridecommandlockouts
\usepackage[
top    = 1 cm,
bottom = 2.50cm,
left   = 1.6cm,
right  = 1.6cm]{geometry}
\usepackage{cite}
\usepackage{amsmath,amssymb,amsfonts}
\usepackage{algorithmic}
\usepackage{graphicx}
\usepackage{textcomp}
\usepackage[table,xcdraw]{xcolor}
\usepackage{romanbar}
\usepackage{tikz}
\usepackage{mathdots}
\usepackage{yhmath}
\usepackage{cancel}
\usepackage{color}
\usepackage{siunitx}
\usepackage{array}
\usepackage{multirow}
\usepackage{gensymb}
\usepackage{tabularx}
\usepackage{booktabs}
\usepackage{verbatim}
\usepackage[ruled,vlined]{algorithm2e}
\usepackage{balance}
\usepackage{multirow}
\usepackage{adjustbox}
\usepackage{comment}

\usetikzlibrary{fadings}
\def\BibTeX{{\rm B\kern-.05em{\sc i\kern-.025em b}\kern-.08em
		T\kern-.1667em\lower.7ex\hbox{E}\kern-.125emX}}
\pdfpagewidth=\paperwidth
\pdfpageheight=\paperheight

\begin{document}
	
	\title{Fuzzy-Logic Based IDS for Detecting Jamming Attacks in Wireless Mesh IoT Networks
		\thanks{This research is part of a project that has received funding from the European Union's Horizon 2020 research and innovation programme under grant agreement Nº739578 and the government of the Republic of Cyprus through the Directorate General for European Programmes, Coordination and Development.}
	}
	
	\author{\IEEEauthorblockN{Michael Savva\authorrefmark{1}, Iacovos Ioannou\authorrefmark{1}\authorrefmark{2}, and Vasos Vassiliou\authorrefmark{1}\authorrefmark{2}}
		\IEEEauthorblockA{\authorrefmark{1}Department of Computer Science, University of Cyprus\\
			\authorrefmark{2}CYENS Centre of Excellence, Nicosia, Cyprus
		}
	}
	
	\maketitle

	\begin{abstract}
		The investigation in this paper targets the design and the evaluation of jamming intrusion detection based on Fuzzy Logic in wireless mesh IoT Networks in a distributed manner. Our approach uses information collected at local nodes and from the sink as input to the fuzzy logic controller. In order to find the best set of inputs, distributed or centralized, we made a comparison between five different combinations of parameters. The investigation uses the values of the ETX, Retransmissions, Packets Drop per terminal (PDPT) and Packet Delivery Ratio (PDR) as inputs to a fuzzy inference system to get Jamming Index (JI) as the system’s output. The proposed method was evaluated based on the following metrics: Accuracy, Precision,  Specificity, False positive rate (FPR),  Recall, False negative rate (FNR) and ROC curve. In order to evaluate this approach, we implement experiments in various scenarios using the Contiki OS and the Cooja simulator tool.
		
	\end{abstract}
	
	\begin{IEEEkeywords}
		Jamming, Intrusion Detection, Fuzzy logic, Internet of Things, Wireless Sensor Networks
	\end{IEEEkeywords}
	
	\section{Introduction}
	\label{intro}
	
	Internet of Things (IoT) comprises sensors and actuators included in embedded devices with the use of IPv4 and IPv6 technologies, enabling these electronic devices to connect and exchange data that can inter-operate within the existing Internet infrastructure. The IoT devices collect valuable data with the help of various technologies and then, autonomously, transfer these data between other devices \cite{hendricks2015trouble}. 
	However, IoT devices and networks are often connected to critical infrastructure networks. As a result, security and protection are needed. IoT networks such as WSNs are vulnerable to several threats as the nature of such devices has many constraints, for example, limited storage, low computation power and limited power consumption \cite{surendar2016indres}. Additionally, the coordination of diverse technologies, heterogeneity and the distributed nature of the network magnifies the threats to the loT system \cite{surendar2016indres}. Due to the constraints involved in the IoT, not all existing security mechanisms can be applied to protect such devices. 
	As a consequence, IoT devices remain exposed to various vulnerabilities, which keep the associated infrastructure and applications in danger \cite{lopez2019extensive}. 
	
	There are so many attacks against IoT networks that can downgrade their performance and functionalities \cite{deogirikar2017security}. However, one of the most common and dangerous attacks is a jamming attack \cite{Jaitly2017}.  
	IoT systems based on wireless mesh technologies are vulnerable to jamming attacks which could drain the battery of target devices, by disrupting their data transmission and making them repeatedly re-transmit \cite{namvar2016jamming}. Jamming in wireless networks has become a significant research problem due to the ease of its execution \cite{grover2014jamming}. Furthermore, jamming threats can only be prevented at the physical layer (PHY) but not at the MAC or network layer. When a wireless network suffers from jamming attacks, its wireless signals are typically overwhelmed by irregular or sophisticated radio jamming signals, making it hard for legitimate wireless devices to decode data packets. Consequently, any approaches at the MAC layer or above are incapable of preventing jamming threats, and innovative anti-jamming approaches are required at the physical layer \cite{pirayesh2021jamming}.
	
	A practical solution for detecting malicious behaviour within the network is an Intrusion Detection System. 
	In this paper, we provide an intelligent and adaptive intrusion detection technique using Fuzzy logic algorithms in a distributed\footnote{The decision is executed at the Node.} and centralised\footnote{The decision is executed at the Sink.} manner, capable of performing detection locally (with the use of ETX, Retransmissions and PDPT as inputs) or centrally (with the help of PDR input) in a network. The main contributions of the paper relate to : i) method to perform local detection; ii) detection of Jamming attacks at the MAC and Network layer; iii) comparison of five different combinations of parameters as inputs to the Fuzzy controller, and iv) evaluation of different jammer settings (48 different locations of the jammer).
	
	The rest of the paper is structured as follows. Section \ref{baground_knowledge_related_work} provides background information on jamming attacks. It also provides related work on approaches addressing jamming attacks using fuzzy logic. The problem description is provided in Section \ref{problem_formulation_and_approaches_specifics}. Specifically, the examined approaches of Anomaly Detection using Fuzzy Logic are provided in Section \ref{fuzzattack}. The efficiency of the investigated approaches is examined, evaluated and compared in Section \ref{performanceevaluation}. Finally, Section \ref{conclussions_and_future_work} includes concluding remarks. 
	
	\section{Background Knowledge and Related Work}
	\label{baground_knowledge_related_work}
	This section provides a brief description of Jamming attacks and related work on using Fuzzy Logic algorithms for detecting jamming attacks.
	
	\subsection{Jamming Attacks}\label{CD2D}
	This subsection provides background knowledge regarding the jamming attack.
	
	Jamming is a type of attack that interferes with the radio frequencies the network nodes are using \cite{wang2006survey}. A jamming source may either be powerful enough to disrupt the entire network or less powerful and only disrupt a smaller portion of the network.
	The most common and dangerous attack which can be proved harmful for wireless mesh WSN or IoT networks is jamming attack \cite{Jaitly2017}. Wireless networks are especially vulnerable to radio jamming attacks for the reason that the jamming attacks are straightforward to launch \cite{pirayesh2021jamming}. An attacker can easily Generate a jamming attack without requiring any special hardware \cite{cordero2015jamming} and without requiring information about the internals of the control system \cite{vanhoef2014advanced}.
	
	\subsubsection{Types of Jamming Attacks}
	According to the literature \cite{xu2005feasibility, Mpitziopoulos2009} there are three types of jammers, which are the most common. The Proactive, the Reactive, and the Specific-Function jammer.
	
	\paragraph{Proactive Jammers}
	A Proactive Jammer attacks the network regardless of any data communication. It randomly transmits bits on the network, making all the functional nodes non-responsive. Finally, it functions only on a single channel and operates until its energy is depleted \cite{Jaitly2017, grover2014jamming}. Proactive Jammers can be categorized into Constant Jammers, Deceptive Jammers and Random Jammers.
	
	\begin{enumerate}
		\item \textbf{Constant Jammer}: A Constant Jammer continuously emits radio signals that interfere with the transmission of the network \cite{vadlamani2016jamming}. Furthermore, the Constant Jammer emits random signals which do not follow any underlying MAC protocol \cite{Mpitziopoulos2009}.
		This type of jammer aims at keeping the channel busy and damaging the nodes' communication \cite{Mpitziopoulos2009}. On the other hand, constant jammer attacks are energy inefficient and also can be easily detected \cite{grover2014jamming}. Moreover, this type of attack can be easily implemented, easily identified and works on single-channel \cite{Jaitly2017}.
		
		\item \textbf{Deceptive Jammer}: Compared to a Constant Jammer, the Deceptive Jammer is more challenging to detect because it transmits legitimate packets instead of random bits. Similarly to the Constant Jammer, the Deceptive Jammer is also energy inefficient due to the continuous transmission; however, it has very easy implementation \cite{grover2014jamming} and difficult detection \cite{Jaitly2017}.
		
		\item \textbf{Random Jammer}: On the other hand, the random jammer alternates from the sleeping mode to the Jamming mode \cite{grover2014jamming, babar2015security}. It can either behave like a Constant Jammer or a Deceptive Jammer during its jamming phase. In contrast to the previous two jammers, this one reduces power consumption \cite{grover2014jamming}. However, it is less effective than the two abovementioned jammers and incapable of jamming during sleep mode\cite{Jaitly2017}.
		
	\end{enumerate}
	
	\paragraph{Reactive Jammers}
	A Reactive Jammer listens to the channel activity. If it identifies an action, it immediately sends out a random signal to collide with the existing signal on the channel \cite{Mpitziopoulos2009}.
	The amount of power required to listen to a channel is much less in comparison with the power needed for proactive jamming \cite{vadlamani2016jamming}. Unfortunately, reactive jammers are difficult to detect, challenging to design, energy inefficient and work on a single-channel \cite{Jaitly2017}.
	
	\begin{enumerate}
		\item \textbf{Reactive RTS/CTS Jammer}: In this attack, the jammer begins the offensive when it senses a request-to-send (RTS) message to transmit from the sender. As a result, the receiver cannot send back a clear-to-send (CTS) reply because the RTS packet sent from the sender was destroyed. Finally, the sender will not send data because it believes that the receiver is busy with another ongoing transmission. Consequently, the sender is not sending data, and the receiver always waiting for the data packet \cite{grover2014jamming}. 
		
		\item \textbf{Reactive Data Acknowledgement Jammer}: In the Data/ACK attack, the jammer destroys the transmissions of data or acknowledgement (ACK) packets. The attacker does not respond until a data transmission begins at the transmitter end. This type of jammer can corrupt data packets or ACK packets. As a result, we can view an increase in retransmissions. That is happening because the data packets are not received correctly at the receiver in case of data transmissions. In the case of ACK, since the sender does not receive the ACKs, it believes something is going wrong at the receiver side \cite{grover2014jamming}. 
		
	\end{enumerate}
	
	\paragraph{Specific-Function Jammer} These types of jammers are manufactured to achieve a specific function. For example, they can be used to interfere with a single channel, or they can cause the jamming of the whole network depending upon their purpose, which means they can minimize their energy consumption or can increase their maximum throughput \cite{Jaitly2017}.

	\begin{enumerate}
		\item \textbf{Follow-on Jammer}: This category of the jammer hops over all available channels very frequently and jams each channel for a short period \cite{Mpitziopoulos2009}. If a transmitter detects an attack at a specific frequency and hops to another frequency, then the follow-on jammer will scan the channel and hop in the spectrum where there is traffic, or they can randomly hop and jam in different frequencies. To conclude, the follow-on jammer is particularly effective against anti-jamming techniques such as the frequency hopping spread spectrum (FHSS), which uses a slow-hopping rate \cite{Jaitly2017, grover2014jamming}.
		
		\item \textbf{Channel-hopping Jammer}: In the Channel-hopping attack, the jammer interferes while hopping between different channels. Besides, a jammer has direct access to channels because it can override the CSMA algorithm of the MAC protocol. Furthermore, Channel-hopping jammers can jam multiple channels at the same time. Therefore, the jammer is quiet and invisible to its neighbours during the discovery phase, and it starts performing attacks on different channels at different times according to a predetermined pseudorandom sequence \cite {Jaitly2017, grover2014jamming}.
		
		\item \textbf{Pulse Noise Jammer}: This attack category can switch channels and jam on different bandwidths at different periods. Moreover, the pulse noise jammer saves energy by turning off and following the jammer's programming. Pulse noise jammers can attack simultaneously in multiple channels \cite{Jaitly2017, grover2014jamming}.
		
	\end{enumerate}

	\subsubsection{Parameters for Jamming Attack Detection} 
	
	The jamming detection parameters that were applied in existing systems will be discussed in this section.
	
	\begin{itemize}
		\item \textbf{Packet Send Ratio (PSR)}: Xu et al. \cite{xu2005feasibility} defined PSR as the ratio of packets that are successfully sent out by a legitimate traffic source compared to the number of packets it intends to send out at the MAC layer.
		
		\item \textbf{Packet Delivery Ratio (PDR)}: Xu et al. \cite{xu2005feasibility} describe PDR as the ratio of packets that are successfully delivered to a destination compared to the number of packets that have been sent out by the sender. \cite{manju2012detection}. From the outcomes, it was recognised that the PSR and PDR were difficult to decide about jamming and its types \cite{vijayakumar2018fuzzy}. 
		
		\item \textbf{Average number of required transmissions per packet (ATX)}:	Heo et al. \cite{heo2017dodge} define the ATX metric as a total number of transmissions divided by the number of successfully received unique packets.
		
		\item \textbf{Number of hops for received packets}: A Hop count referes to the number of routers that a packet goes through from its source to its destination \cite{agah2007preventing}.	
		
		\item \textbf{Throughput}: Agah and Das \cite{agah2007preventing} define Throughput as the measure that characterises the total number of forwarded packets over the total number of received packets.
		
		\item \textbf{Bit Error Rate (BER)}: According to Strasser et al. \cite{strasser2010detection}, the BER is calculated as the ratio of the number of corrupted bits to the number of total bits received by a node during a transmission session. However, it is hard to measure the BER by a sensor node since it needs to collect a tremendous amount of data. Moreover, this method cannot classify different kinds of jamming attacks \cite{vijayakumar2018fuzzy}.
		
		\item \textbf{Packets Dropped per Terminal (PDPT)}: Misra et al. \cite{Misra2010}, Cakiroglu et al. \cite{2008jamming} and Balarengadurai et al. \cite{balarengadurai2012fuzzy} used PDPT for detection of Jamming attacks. PDPT refers to the ratio of the number of received packets that have not passed the Cyclic Redundancy Check (CRC) carried out by the node, to the total number of packets received by the node over a given period.
		
		\item \textbf{Signal-to-Noise Ratio (SNR)}: SNR measured as the ratio of the received signal power at a node to the received noise power at the node. Misra et al. \cite{Misra2010}, Balarengadurai et al. \cite{balarengadurai2012fuzzy} and Sasikala and Rengarajan \cite{sasikala2015intelligent} used SNR as a metric to detect Jamming attacks. The SNR is a useful metric to identify the behaviour of jamming at the physical layer \cite{vijayakumar2018fuzzy}.
		
		\item \textbf{Energy Consumption}: Cakiroglu et al. \cite{2008jamming} define Energy Consumption as the approximated energy amount consumed in a specified time for a sensor network \cite{manju2012detection, babar2015security}.
		
		\item \textbf{Delay}: Delay in \cite{babar2015security} is calculated as a total time from the transmission of a packet from a node to the sink, to the time when the sink received the packet.
		
		\item \textbf{Received Signal Strength Indicator (RSSI)}: RSSI \cite{manju2012detection, Meenalochani2019, vijayakumar2018fuzzy} is defined as the power content of the radio signal when received by the receiver. 
		
		\item \textbf{Packet Loss Ratio (PLR)}: PLR is calculated as the number of packets lost divided by the number of packets sent \cite{Meenalochani2019, chen2018analysis}.
		
		\item \textbf{Routing Overhead}: Chen et al. \cite{chen2018analysis} proposed routing overhead as the average number of routing packets (including DIS, DIO, and DAO packets) transmitted in the whole network every minute.
		
		\item \textbf{Expected Transmission Count (ETX)}: ETX \cite{renofio2016dynamics} represents the expected number of transmissions required to successfully transmit and acknowledge a packet on a wireless link.
		
	\end{itemize}
	
	Our solution uses as parameters in the fuzzy logic the values of the ETX, Retransmissions, PDPT and PDR as inputs to a fuzzy inference system. The decision of choosing these values
	is based on selecting matrics where collected at the node or Sink node (PDR). Additionally, from our empirical experience and also from our experimental results, these values extremely increase when having a Jammer attack. 
	
	\subsection{Related Work on Using Fuzzy Logic Algorithms for Detecting Jamming Attacks}
	\label{RTS}
	
	Fuzzy logic dealing with vagueness and imprecision has a capacity to describe imprecise forms of reasoning in areas where firm decisions have to be made in indefinite conditions and is found to be proper for intrusion detection \cite{bhattacharjee2013fuzzy}.
	Following the literature, Detection frameworks based on fuzzy logic have the capability to calculate ambiguous information availability \cite{sherazi2019ddos}. Fuzzy logic can be making actual-time decisions, even with incomplete knowledge. Conventional control systems rely on an accurate representation of the environment, which commonly does not exist in reality. Fuzzy logic systems, which can handle the linguistic rules naturally, are suitable in this respect. Moreover, it can be used for context by blending different parameters – rules combined to produce a suitable result \cite{ balarengadurai2012fuzzy}. Furthermore, Fuzzy rules leave us to efficiently and easily construct if-then rules that reflect general ways of describing security attacks \cite{dickerson2000fuzzy}. Thus, fuzzy logic can be an adequate means of defining network attacks \cite{dickerson2000fuzzy}.
	AI methods such as decision trees, neural networks and fuzzy logic are applied for detecting anomalies in a network, in which a fuzzy-based system presents important advantages over other AI techniques \cite{shanmugavadivu2011network}.
	Our approach uses a combination of the following metrics ETX, Retransmissions, PDPT and PDR  as inputs to a Fuzzy inference system (Mamdani’s Fuzzy Inference System)\cite{mamdani1999experiment} to get Jamming Index (JI) as the output of the system. The JI value is between 0 and 1, signifying No Jamming to Absolute Jamming, respectively.
	According to the literature, the following works take into account the jamming attack and use Fuzzy logic algorithms to detect malicious activity in wireless mesh networks.
	
	Misra et al. \cite{Misra2010} proposed a fuzzy inference system for Jamming attack detection. In this approach, network nodes receive input values, while the base station does the jamming detection following a centralized methodology. The nodes send three inputs to the base station: the number of total packets received during a specified period, the number of packets dropped during the period, and the received signal strength (RSS). With these metrics, the base station can calculate the PDPT and SNR. Afterwards, the central node uses the values of PDPT and SNR as inputs to the fuzzy inference system to extract the jamming index. Finally, they make a confirmation of a Jamming Attack on a node. The validation is done through a 2-Means Clustering algorithm that constructs a confirmatory check through the study of the neighborhood of a node, to ascertain the correctness of the JI grade allotted to that node compared to the JI distributed to its neighbour nodes. This work is done using NS2, MATLAB and Simulink simulator and exanimated four types of jammers constant, deceptive, random, and reactive in 720 different simulations setups. In simulations, there are four positions for the jammer, two inside and two outside the grid,  Six sets of inter-nodal distances: 5, 10, 15, 20, 25, and 30 meters and three sets of nodes: 25, 50, and 100. Finally, researchers have done simulations using standard power at the jammer node and high energy at the jammer node.
	
	Balarengadurai and Saraswathi \cite{balarengadurai2012detection} detect Jamming attacks at the PHY and MAC layers in IEEE 802.15.4 low rate wireless personal area network using Fuzzy logic systems. This system used three inputs being sent by the nodes to the base station. The inputs are a) the number of total packets received by it during a specified period, b) the number of packets dropped by it during the period, and c) the received signal strength (RSS). Afterwards, the base station calculates the PDPT and SNR from these values, and then the base station uses the metrics of PDPT and SNR as inputs to a fuzzy logic system to get Jamming Index as the output of the system. Finally, the confirmation of the Jamming attack Detection is done through the Fuzzy K-Means Clustering. In these experiments, used S-MAC protocol as the MAC standard. In simulations, the nodes sent one packet every 5 seconds for light traffic and two packages every 1 second for heavy traffic. In addition, the authors examined four types of jammers constant, deceptive, random, and reactive—the evaluation of this approach simulated in a Network Simulation environment.
	
	Vijayakumar et al. \cite{vijayakumar2018fuzzy} proposed a fuzzy logic-based jamming detection algorithm (FLJDA) to detect the presence of jamming in downstream data communication for cluster‐based wireless sensor networks. FLJDA monitors the behaviour of nodes by computing the jamming probability using two inputs in a fuzzy logic system: the packet delivery ratio and the received signal strength indicator. The evaluation of this approach is simulated in MATLAB. The authors examined four types of jammers constant, deceptive, random, and reactive.
	
	Meenalochani and Sudha \cite{Meenalochani2019} proposed a hybrid algorithm based on Fuzzy logic and Ant Colony Optimization for the detection of jamming attacks. The evaluation of this approach is simulated in MATLAB. The authors used fuzzy logic to detect the interference node and the ant colony to route the data even in the presence of jamming. The ant colony approach discards the congested node and identifies a path from source to destination for successful transmission. The fuzzy logic system used PDR, PLR, and RSSI as input to determine the node's jamming percentage. The authors used 12 wireless Zigbee real nodes where node one was assigned as Base Station and connected to a Laptop to display the received data in real-time.
	
	In contrast with existing solutions shown above, which also detect jamming attacks using a fuzzy logic algorithm, our approach uses only two metrics as input. It acts with a lightweight edition of an IDS to achieve a high-performance evaluation system. Our solution uses only one algorithm in comparison with Mirsa et al. \cite{Misra2010} who used the 2-means clustering algorithm additionally in order to achieve the best results. Similarly, Balarengadurai and Saraswathi \cite{balarengadurai2012detection} used an additional K-means clustering algorithm. Finally, Meenalochani and Sudha \cite{Meenalochani2019} used three metrics as inputs, the PDR, PLR and RSSI and additionally used the Ant Colony Optimization algorithm. Overall, none of the researchers examined so comprehensively the position of the jammer. Our approach examines forty-eight different positions of the jammers with equivalent simulations in a distributed and centralised manner. More precisely, the identification of the jammer is executed at each node (with ETX, Retransmissions and PDPT parameters) and at sink (with PDR parameter). 
	Moreover, all approaches use input metrics from the physical layer and make the detection decision centralised, unlike our solution where the input values are Network layer metrics and make the detection decision locally.
	
	\section{Problem Description and Detection Methodology}
	\label{problem_formulation_and_approaches_specifics}
	This paper tries to tackle the problem of identifying jamming attacks using Fuzzy logic algorithms in IoT networks. We implement experiments in various scenarios using the Contiki OS and the Cooja simulator tool. In our approach, node and network information was collected from the local nodes and from the sink and used as input to the fuzzy logic controller to implement jammer detection. Our work uses the following local metrics: the ETX, Retransmissions, PDPT and the PDR (central metric) as inputs to a fuzzy inference system to get Jamming Index (JI) as an output value. The input parameters examined for the fuzzy logic are selected because these values are network layer metrics and at the same time they can be collected and processed distributed at the node. The main purpose of this paper is to find the best input parameters for our Fuzzy controller. In order to achieve this, we made a comparison between five different combinations of inputs. The proposed method was evaluated based on the Accuracy, Precision,  Specificity, FPR,  Recall, FNR and ROC curve.
	
	\subsection{Approaches of Anomaly Detection using Fuzzy logic}\label{fuzzattack}
	
	In this paper we examine five different combinations of the four metrics. Firstly, we get the results where the inputs of the Fuzzy controller are ETX and Retransmissions. Secondly, we used the combination of PDPT and Retransmissions. Thirdly, we used the combination of PDR and Retransmissions. Fourth, we used the combination of ETX and PDR. Finally, we used the combination of ETX and Drop. The combination of PDR and PDPT is not taken into consideration for the reason the two inputs are dependent variables.
	
	We define three fuzzy sets each over the four universes of discourse (inputs), ETX, Retransmissions, PDPT and PDR : LOW, MEDIUM, and HIGH. In addition, we define four fuzzy sets over the universe of discourse (output), JI: NO attack, LOW, MEDIUM, and HIGH. We use Mamdani's model \cite{mamdani1999experiment}, where combinations of ETX, Retransmissions, PDPT and PDR are the crisp inputs to the system, and JI is the crisp output obtained from the system after defuzzification using the centroid method.
	
	
	As generated through the Cooja simulator, multiple sets of four crisp inputs, ETX, Retransmissions, PDPT and PDR, are first planned into fuzzy membership functions.
	
	A trapezoid shape is preferred to define fuzzy membership functions because of two reasons: firstly, it can be mathematically manipulated to be very close to the most natural feature, the Gaussian or Bell function, and secondly, it can be easily manipulated to be an unsymmetrical function (as required in the instant case) where the same cannot be done so easily with the Gaussian or Bell functions\cite{Misra2010}.
	In this study, we chose trapezoidal shapes as an appropriate membership function for our fuzzy logic controller. The decision of which of the methods is going to use depends completely on the problem size and problem type \cite{sadollah2018introductory}. The choice of trapezoidal shapes depends on the distribution of our data. In comparison with Gaussian fuzzy sets, the Trapezoidal shapes are easy to implement and fast to calculate \cite{sadollah2018introductory}. Following the literature \cite{wu2012twelve}, there are two strategies for constructing the fuzzy sets: a) model-driven and b) knowledge-driven. Gaussian fuzzy sets can only be created from the model-driven approach, whereas trapezoidal fuzzy sets can be constructed from both model-driven and knowledge-driven approaches. As a result, working with the trapezoidal fuzzy sets gives the user more freedom in membership function construction \cite{wu2012twelve}. To conclude, Trapezoidal fuzzy logic controllers are simpler in analysis \cite{wu2012twelve}. 
	
	\subsection{Membership Functions}\label{memfunc}
	We designate the membership functions below:
	
	\begin{equation}
		\mu_{set} (uod) = \left \{
		\begin{aligned}
			&\frac{uod-\alpha}{b-\alpha}, && \alpha \leq uod \leq b \\
			&1, && b < uod < c  \\
			&\frac{d-uod}{d-c}, && c \leq uod \leq d \\
			&0, && \text{otherwise}
		\end{aligned} \right.
	\end{equation}

\begin{figure}[h!]
		\centering
		\includegraphics[width=0.80\linewidth]{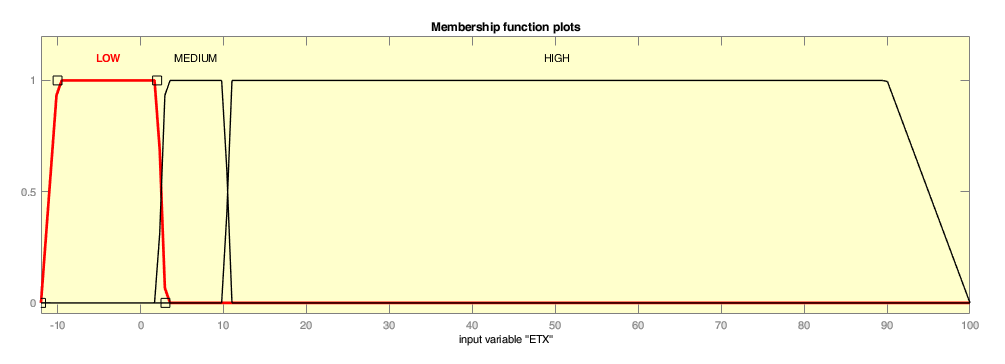}
		\caption{The trapezoidal Membership function plots for the input ETX.}
		\label{Figures_ETX_member}	
	\end{figure}

\begin{figure}[h!]
		\centering
		\includegraphics[width=0.80\linewidth]{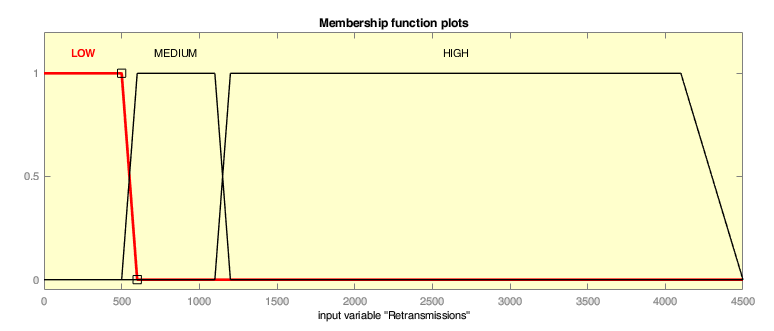}
		\caption{The trapezoidal Membership function plots for the Retransmissions.}\
		\label{Figures_retransm_member}
	\end{figure}

\begin{figure}[h!]
		\centering
		\includegraphics[width=0.80\linewidth]{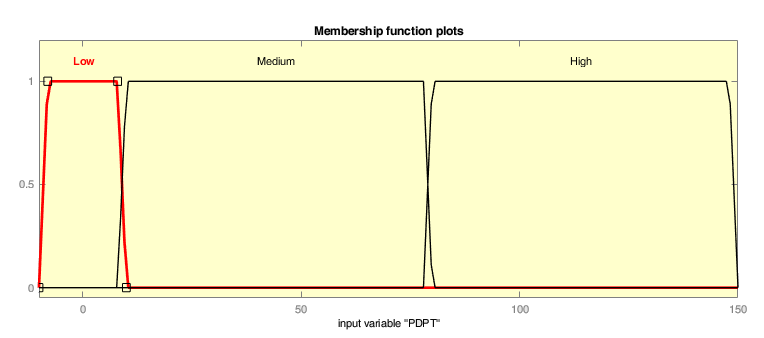}
		\caption{The trapezoidal Membership function plots for the input PDPT.}
		\label{Figures_Drop_REtrans_PDPT}	
	\end{figure}

\begin{figure}[h!]
		\centering
		\includegraphics[width=0.80\linewidth]{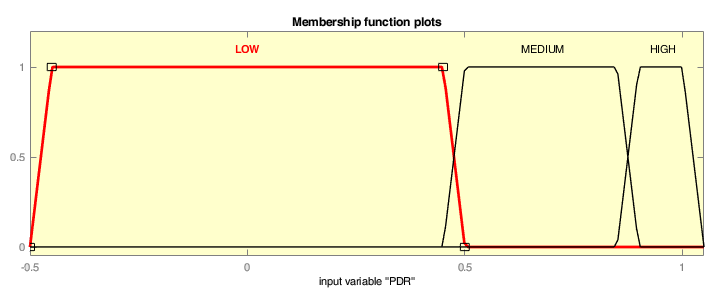}
		\caption{The trapezoidal Membership function plots for the input PDR.}\
		\label{Figures_PDR_retrans_PDR}	
	\end{figure}

\begin{figure}[h!]
		\centering
		\includegraphics[width=0.80\linewidth]{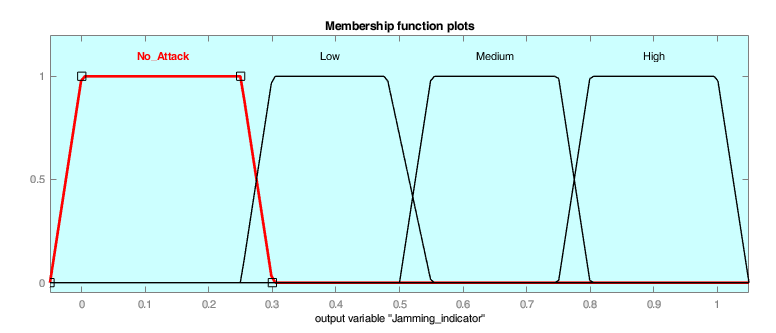}
		\caption{The trapezoidal Membership function plots for the output, Jamming index (JI).}
		\label{Figures_JI_member}
	\end{figure}

\begin{figure} 
		\centering
		\includegraphics[width=0.80\linewidth]{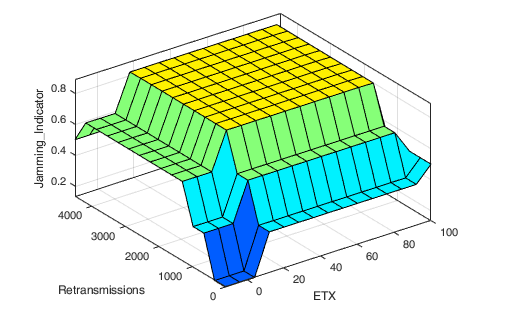}
		\caption{Input-output surface corresponding to the membership values of inputs (ETX,Retransmissions) and output (JI).}
		\label{Surface_2inputs}
	\end{figure}

\begin{figure} 
		\centering
		\includegraphics[width=0.80\linewidth]{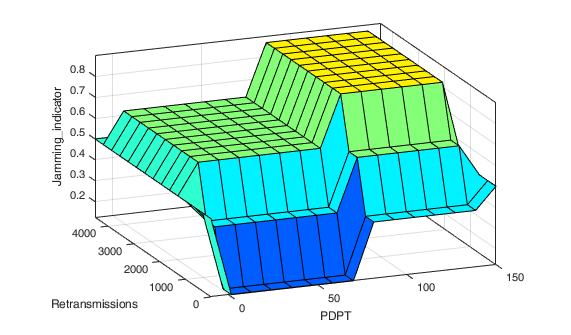}
		\caption{Input-output surface corresponding to the membership values of inputs (PDPT,Retransmissions) and output (JI).}
		\label{Surface_PDPT_Retransmissions}
	\end{figure}

\begin{figure} 
		\centering
		\includegraphics[width=0.80\linewidth]{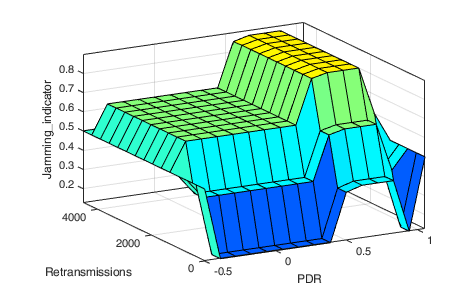}
		\caption{Input-output surface corresponding to the membership values of inputs (PDR,Retransmissions) and output (JI).}
		\label{Surface_PDR_Retransmissions}
	\end{figure}

\begin{figure} 
		\centering
		\includegraphics[width=0.80\linewidth]{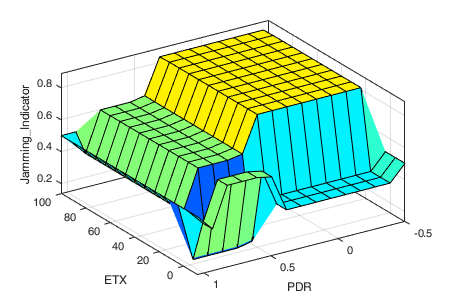}
		\caption{Input-output surface corresponding to the membership values of inputs ETX,PDR) and output (JI).}
		\label{Surface_ETX_PDR}
	\end{figure}

\begin{figure} 
		\centering
		\includegraphics[width=0.80\linewidth]{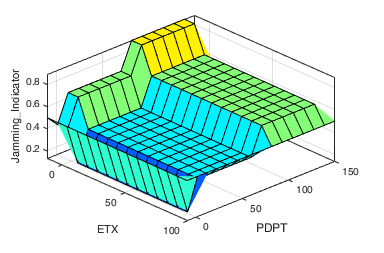}
		\caption{Input-output surface corresponding to the membership values of inputs ETX, PDPT) and output (JI).}
		\label{Surface_ETX_PDPT}
	\end{figure}

	Where the separate values of the variables are as provided in Table \ref{table_ETX_Retrans}. The values of the variables, as presented in Table \ref{table_ETX_Retrans}, have been fixed through by the improvement of these values through a feedback factor generated by contrasting the original result (the output, JI of the method) and the expected outcome (the JI value).
	A fuzzy logic system can be constructed using observed data \cite{Meenalochani2019}.
	In order to find the ranges of the variables, we observe the values with the minimum value, the maximum value, the average value and the standard deviation. Furthermore, we observe the distribution of the values to build the membership functions.
	The graphical illustrations of these trapezoidal functions in respect of ETX, Retransmissions, PDPT, PDR and JI are shown in Figures \ref{Figures_ETX_member}, \ref{Figures_retransm_member}, 	\ref{Figures_Drop_REtrans_PDPT}, 	\ref{Figures_PDR_retrans_PDR}	 	and \ref{Figures_JI_member}, respectively. Additionally, in Figures \ref{Surface_2inputs}, \ref{Surface_PDPT_Retransmissions}, \ref{Surface_PDR_Retransmissions}, \ref{Surface_ETX_PDR} and \ref{Surface_ETX_PDPT}  we show the Input-output surface corresponding to the membership values of inputs.
	
	\begin{table}
		\centering
		\caption{Values of variables used in the definition of membership functions}
		\label{table_ETX_Retrans}
		\resizebox{\linewidth}{!}{%
			\begin{tabular}{|l|l|l|l|l|l|}
				\hline
				\textbf{Universe of discourse (uod)}    & Set       & a & b & c & d \\ \hline
				\multirow{3}{*}{ETX}                           & LOW       & -4 &-2   &2   &3   \\ \cline{2-6}
				& MEDIUM    &2   &3   &10   &11   \\ \cline{2-6}
				& HIGH      & 10  &11   &18   &22   \\ \hline
				\multirow{3}{*}{Retransmissions}                            & LOW       &-150  &-100   &500   &600   \\ \cline{2-6}
				& MEDIUM    & 500  &600   &1100   &1200   \\ \cline{2-6}
				& HIGH      &  1100 & 1200 &  4100 &4500   \\ \hline
				\multirow{3}{*}{PDPT}                           & LOW       & -10 &-8   &8   &10   \\ \cline{2-6}
				& MEDIUM    &8   &10   &78   &80   \\ \cline{2-6}
				& HIGH      & 78  &80   &148   &150   \\ \hline
				\multirow{3}{*}{PDR}                            & LOW       &-0.5  &0   &0.45   &0.5   \\ \cline{2-6}
				& MEDIUM    & 0.45  &0.5   &0.85   &0.9   \\ \cline{2-6}
				& HIGH      &  0.85 & 0.9 &  1 &1.05   \\ \hline
				\multirow{4}{*}{Jamming Indicator (JI)} & NO ATTACK & -0.05  &0   &0.25   &0.3   \\ \cline{2-6}
				& LOW       &  0.25 & 0.3  &0.5   & 0.55  \\ \cline{2-6}
				& MEDIUM    & 0.5  &0.55   &0.75   &0.8   \\ \cline{2-6}
				& HIGH      &  0.75 &0.8   &1   &1.05   \\ \hline
			\end{tabular}
		}
	\end{table}

	\subsection{Detection Algorithm}\label{DetectAlgo}
	The algorithm for detecting the jamming node is shown below.
	
	\begin{algorithm}
		\begin{algorithmic}
			\caption{Algorithm for detection of jammed node using fuzzy inference system}
			\STATE Input:
			\begin{itemize}
				\item ETX and Retransmissions (Distributed), PDPT and Retransmissions (Distributed), PDR and Retransmissions (Centralized), ETX and PDR (Centralized)
			\end{itemize}
			
			\STATE Output:
			\begin{itemize}
				\item Jamming Indicator (JI)
				
			\end{itemize}
			\STATE BEGIN
			\begin{itemize}
				\item Get the values of ETX and Retransmissions, PDPT and Retransmissions, PDR and Retransmissions, ETX and PDR of the nodes
				\item Fuzzify the crisp input parameters: ETX and Retransmissions, PDPT and Retransmissions, PDR and Retransmissions, ETX and PDR
				\item Apply the fuzzy rule base and get the fuzzy output
				\item Defuzzify the fuzzy output to get the crisp output
				\item The defuzzified crisp output gives the percentage of jamming of a node
			\end{itemize}
			
			\STATE END
		\end{algorithmic}
	\end{algorithm}
	
	In the approach where the inputs are ETX and Retransmissions the fuzzy rule base is given below:
	\begin{enumerate}
		\scriptsize
		\item If ETX is LOW and Retransmissions is LOW then JI is No ATTACK.
		\item If ETX is LOW and Retransmissions is MEDIUM then JI is LOW.
		\item If ETX is LOW and Retransmissions is HIGH then JI is MEDIUM.
		\item If ETX is MEDIUM and Retransmissions is LOW then JI is No ATTACK.
		\item If ETX is MEDIUM and Retransmissions is MEDIUM then JI is LOW.
		\item If ETX is MEDIUM and Retransmissions is HIGH then JI is MEDIUM.
		\item If ETX is HIGH and Retransmissions is LOW then JI is LOW.
		\item If ETX is HIGH and Retransmissions is MEDIUM then JI is MEDIUM.
		\item If ETX is HIGH and Retransmissions is HIGH then JI is HIGH.
	\end{enumerate}
	
	In the approach where the inputs are PDPT and Retransmissions, the fuzzy rule base is given below:
	\begin{enumerate}
		\scriptsize
		\item If PDPT is LOW and Retransmissions is LOW then JI is No ATTACK.
		\item If PDPT is LOW and Retransmissions is MEDIUM then JI is LOW.
		\item If PDPT is LOW and Retransmissions is HIGH then JI is MEDIUM.
		\item If PDPT is MEDIUM and Retransmissions is LOW then JI is No ATTACK.
		\item If PDPT is MEDIUM and Retransmissions is MEDIUM then JI is LOW.
		\item If PDPT is MEDIUM and Retransmissions is HIGH then JI is MEDIUM.
		\item If PDPT is HIGH and Retransmissions is LOW then JI is LOW.
		\item If PDPT is HIGH and Retransmissions is MEDIUM then JI is MEDIUM.
		\item If PDPT is HIGH and Retransmissions is HIGH then JI is HIGH.
	\end{enumerate}
	
	In the approach where the inputs are PDR and Retransmissions, the fuzzy rule base is given below:
	\begin{enumerate}
		\scriptsize
		\item If PDR is LOW and Retransmissions is LOW then JI is No ATTACK.
		\item If PDR is LOW and Retransmissions is MEDIUM then JI is LOW.
		\item If PDR is LOW and Retransmissions is HIGH then JI is MEDIUM.
		\item If PDR is MEDIUM and Retransmissions is LOW then JI is LOW.
		\item If PDR is MEDIUM and Retransmissions is MEDIUM then JI is MEDIUM.
		\item If PDR is MEDIUM and Retransmissions is HIGH then JI is HIGH.
		\item If PDR is HIGH and Retransmissions is LOW then JI is NO ATTACK.
		\item If PDR is HIGH and Retransmissions is MEDIUM then JI is LOW.
		\item If PDR is HIGH and Retransmissions is HIGH then JI is MEDIUM.
	\end{enumerate}

	In the approach where the inputs are ETX and PDR, the fuzzy rule base is given below:
	\begin{enumerate}
		\scriptsize
		\item If ETX is LOW and PDR is LOW then JI is LOW.
		\item If ETX is LOW and PDR is MEDIUM then JI is MEDIUM.
		\item If ETX is LOW and PDR is HIGH then JI is LOW.
		\item If ETX is MEDIUM and PDR is LOW then JI is LOW.
		\item If ETX is MEDIUM and PDR is MEDIUM then JI is NO ATTACK.
		\item If ETX is MEDIUM and PDR is HIGH then JI is NO ATTACK.
		\item If ETX is HIGH and PDR is LOW then JI is HIGH.
		\item If ETX is HIGH and PDR is MEDIUM then JI is MEDIUM.
		\item If ETX is HIGH and PDR is HIGH then JI is LOW.
	\end{enumerate}

	In the approach where the inputs are ETX and PDPT, the fuzzy rule base is given below:
	\begin{enumerate}
		\scriptsize
		\item If ETX is LOW and PDPT is LOW then JI is LOW.
		\item If ETX is LOW and PDPT is MEDIUM then JI is MEDIUM.
		\item If ETX is LOW and PDPT is HIGH then JI is HIGH.
		\item If ETX is MEDIUM and PDPT is LOW then JI is NO ATTACK.
		\item If ETX is MEDIUM and PDPT is MEDIUM then JI is LOW.
		\item If ETX is MEDIUM and PDPT is HIGH then JI is MEDIUM.
		\item If ETX is HIGH and PDPT is LOW then JI is NO ATTACK.
		\item If ETX is HIGH and PDPT is MEDIUM then JI is LOW.
		\item If ETX is HIGH and PDPT is HIGH then JI is MEDIUM.
	\end{enumerate}

	\section{Performance Evaluation}
	\label{performanceevaluation}
	
	\subsection{Methodology}
	
	In order to evaluate this approach, we implement experiments in various scenarios with the use of Contiki OS and the Cooja simulator tool. In our study, we implement a Deceptive Jamming attack using the JamLab \cite{boano2011jamlab} suite, a Contiki-based library that allows repeatable experiments with radio interference. Our jammer continuously emits signals which include legitimate packets that interfere with the communication of the network.
	
	\subsection{Simulation Environment}
	In our study, we made experiments with the jammer placed in predicted scenarios within a grid of 25 nodes. Our implementation was constructed and evaluated via Cooja O/S and Contiki simulator tool. We extracted the simulation data from the Cooja Simulator, and we processed the data in Python and Matlab.
	
	We placed 25 nodes on a grid in an area of 160 * 160 meters including a central node as a Sink. All nodes have a distance of 40 meters from each other. All nodes are equivalent to TelosB nodes and have a 50-meter transmission range and 70-meter interference range. Each node transmits one data packet of 48 bytes every 10 seconds. 
	
	Additionally, we run three different scenarios regarding the position of the sink. The situations are a) sink in the middle of the grid b) sink on top middle of the grid and c) sink on top left edge of the network. Each scenario run 16 different jammer positions and a healthy (benign) scenario. In accordance with the literature, these three scenarios were used in published works \cite{ioannou2020accurate, ioannou2017intrusion, Misra2010}. The experimental parameters of the nodes are shown in Table \ref{Experimental_Parameters_of_Nodes}.
	
	\begin{table}[h!]
		\tiny
		\centering
		\caption{Experimental Parameters of Nodes}\label{Experimental_Parameters_of_Nodes}
		\begin{tabular}{|l|l|}
			\hline
			\textbf{Parameter}      &    \\ \hline
			\textbf{No. of nodes}                                                                      &      25                                                       \\ \hline
			\textbf{Area Size}                                                                      &      160 * 160 meters                                                       \\ \hline
			\textbf{Sensor nodes}                                                                      &                                                           TelosB nodes \\ \hline
			\textbf{Sink Position}                                                                      &      1.	Sink Middle
			2.	Sink Top Middle
			3.	Sink Top left edge
			
			\\ \hline
			\textbf{Scenario duration}                                   &   15 minutes                                                  \\ \hline
			\textbf{Transmission rate}                                   &   Generate 1 packet of 48 bytes per 10 seconds 
			\\ \hline
			\textbf{Propagation Model}  &          Unit Disk Graph                                               \\ \hline
			\textbf{Transmission range}           &   50 m                                                                \\ \hline
			\textbf{Interference range}                 &          70 m                                                                \\ \hline
			\textbf{Routing Protocol}                 &          RPL                                                              \\ \hline
			\textbf{MAC layer}                 &          CSMA                                                              \\ \hline
			\textbf{RDC layer}                 &          ContikiMAC                                                              \\ \hline
			\textbf{Channel Check Rate}                 &          8 Hz                                                              \\ \hline
		\end{tabular}
	\end{table}
	
	We used the Routing Protocol for Low power and Lossy links (RPL). For Medium Access Control (MAC) layer we used the CSMA and for Radio Duty Cycle (RDC) we used ContikiMAC with the Channel Check Rate of the 8 Hz. 
	
	The Jamming node is using nullmac and nullrdc protocols with the Channel Check Rate of the 128 Hz. The experimental parameters of the jammer are shown in Table \ref{Experimental_Parameters_of_Jammer}.
	
	\begin{table}[h!]
		\tiny
		\centering
		\caption{Experimental Parameters of Jammer}\label{Experimental_Parameters_of_Jammer}
		\begin{tabular}{|l|l|}
			\hline
			\textbf{Parameter}      &    \\ \hline
			\textbf{No. of nodes}                                                                      &      1                                                       \\ \hline
			\textbf{Sensor nodes}                                                                      &                                                           TelosB nodes \\ \hline
			\textbf{Type of Jammer}                                                                      &                                                           Deceptive \\ \hline
			\textbf{Transmission range}           &   50 m                                                                \\ \hline
			\textbf{Interference range}                 &          70 m                                                                \\ \hline
			\textbf{MAC layer}                 &          nullmac                                                              \\ \hline
			\textbf{RDC layer}                 &          nullrdc                                                              \\ \hline
			\textbf{Channel Check Rate}                 &          128 Hz                                                              \\ \hline
		\end{tabular}
	\end{table}
	
	\begin{figure} [h!]
		\centering
		\includegraphics[width=0.6\linewidth]{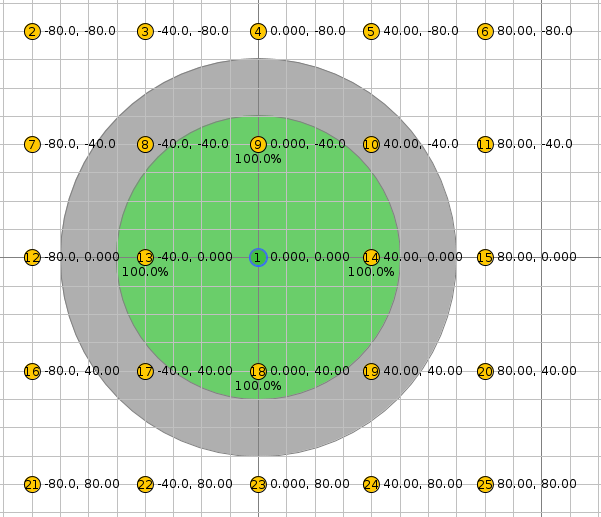}
		\caption{Simulation Set-up and Configuration}\label{SetupConfiguration}
		\label{Figures:5}
	\end{figure}
	
	Figure \ref{SetupConfiguration} shows the simulation set-up and configuration where the Sink is located in the middle of the grid.
	The green area shows the transmission range of a node. More specifically, each node can communicate with all other nodes within the green circle. For instance, the Sink node, which has number 1, can communicate with nodes 9, 13, 14 and 18. Additionally, the grey circle around the green area displays the interference range. For example, when node 1 sends packets, nodes 8, 17, 10 and 19 in the grey area cannot receive packets and they are not able to receive packets from other nodes when the node 1 communicates simultaneously \cite{voigt2009based}.
	
	For each topology, we defined two types of scenarios: a malicious situation in which a compromised node is placed in the network, and a no attack scenario in which all nodes are in healthy condition.
	We run in total 160 malicious scenarios for each topology and one in the reasonable condition where we repeat the normal state ten times to view the deviation of the results. The average costs were used to compare states between normal conditions and attack conditions.
	We choose the grid scenario to do our experiments because its simplicity helps us build the security framework and understand its weaknesses. 
	
	The network topology with the Sink in the middle shows the best case scenario of all three, in which there are four nodes that can be used to reach the Sink and the maximum number of hops from the Sink is four. Additionally, the network topology with the Sink in the top-middle of the grid has three nodes that can directly access the Sink and the maximum number of hops is six. The network topology with the Sink at the top edge indicates the worst-case scenario of all three, has only two nodes that can directly access the Sink and the maximum number of hops is eight \cite{ioannou2020accurate}. We performed the 16 different scenarios and placed a jammer in each one.

\begin{figure} [h!]
		\centering
		\includegraphics[width=0.6\linewidth]{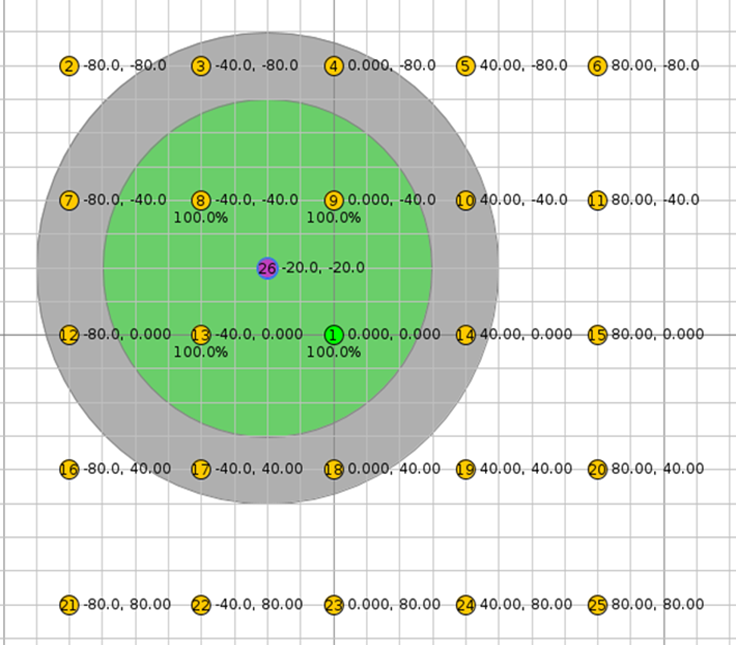}
		\caption{Sink in the Middle of the Grid, Jammer Position 6}\label{Sink_in_the_Middle_Jammer_Position_6}
		\label{Sink in the Middle of the grid Jammer Position 6}
	\end{figure}

	Figure \ref{Sink_in_the_Middle_Jammer_Position_6} shows a jammer in position 6 with coordinates (-20,-20) when the sink is located in the middle of the grid. From the Figure it can be observed that the attacker jams the nodes with number 8, 9, 13, and the sink. Additionally, the malicious node attacks the nodes with numbers 3, 4, 7, 10, 12, 14, 17, and 18.

	\subsection{Jamming attack strategy}
	In this study we used jammers from the JamLab\cite{boano2011jamlab} implementation. More specifically, we used a modified proactive deceptive jammer, with an ON-OFF pattern. The jammer is ON for one second and OFF for 333ms. When ON, the jammer continuously emits interference signals. The attacker sends packets with data that are not recognised by the nodes. The length of the transmitted data sequence is 8 kilobytes. Therefore, the nodes can transfer packets to the other nodes when the intrusion signal is off. The selection of the specific jamming attack is based on the attributes of the jammer, because it is a simpler jammer without a lot of abequites therefore the usage is simple and straightforward. Finally, each malicious node is placed in the centre of four nodes of the network.

	\subsection{Results}
	\label{results}
	
	In this study, we use the Confusion Matrix as shown in Table \ref{conf_matrix} 
	for the description of the performance. 
	
	\begin{table}[h!]
		\small
		\centering
		\caption{Confusion matrix for Jamming attack detection }\label{conf_matrix}
		\begin{tabular}{clll}
			\multicolumn{1}{r}{}                                  &                                         & \multicolumn{2}{c}{\textbf{Detection}}                                         \\ \cline{3-4}
			& \multicolumn{1}{l|}{}                   & \multicolumn{1}{l|}{\textbf{Attack}} & \multicolumn{1}{l|}{\textbf{No Attack}} \\ \cline{2-4}
			\multicolumn{1}{c|}{\multirow{2}{*}{\textbf{Actual}}} & \multicolumn{1}{l|}{\textbf{Attack}}    & \multicolumn{1}{l|}{TP}              & \multicolumn{1}{l|}{FN}                 \\ \cline{2-4}
			\multicolumn{1}{c|}{}                                 & \multicolumn{1}{l|}{\textbf{No Attack}} & \multicolumn{1}{l|}{FP}              & \multicolumn{1}{l|}{TN}                 \\ \cline{2-4}
		\end{tabular}
	\end{table}
	
	where: TP = True Positive, FP = False Positive, TN = True Negative, FN = False Negative \cite{newman2004intrusion}.
	
	The confusion matrix is a matrix that represents true and false classification results \cite{kumar2014evaluation}. From the Confusion Matrix, we calculate the indicators of  Accuracy, Precision,  Specificity, FPR,  Recall and FNR.
	
	The \textbf{True Positive} state is when the IDS identifies an activity as an attack, and the event is an attack. A real positive is a successful identification of an attack \cite{newman2004intrusion}.
	
	The \textbf{True Negative} state is similar. This state is when the IDS identifies an activity as acceptable behaviour, and the activity is acceptable. A true negative is successfully ignoring acceptable behaviour. Neither of these states is harmful as the IDS is performing as expected \cite{newman2004intrusion}.
	
	The \textbf{False Positive} state is when the IDS identifies an activity as an attack, but the action is acceptable behaviour. A false positive is a false alarm \cite{newman2004intrusion}.
	
	The \textbf{False Negative} state is the most severe and dangerous state. This situation is when the IDS identifies an activity as acceptable when the event is an attack \cite{newman2004intrusion}.

	\textbf{Accuracy} defined as the percentage of correctly classified records over the total number of records\cite{javaid2016deep}.
	The equation for Accuracy is shown below.
	
	\begin{equation}\label{accuracy}
		\small
		\begin{aligned}
			Accuracy =&  \frac{TP+TN}{(TP+TN+FP+FN) }
		\end{aligned}
	\end{equation}

	Additionally, \textbf{Precision} is the ratio of correctly predicted positive observations to the total predicted positive observations. Equation \ref{Precision} shows how the Precision is calculated.
	
	\begin{equation}\label{Precision}
		\small
		\begin{aligned}
			Precision =&  \frac{TP}{(TP+FP)}
		\end{aligned}
	\end{equation}
	
	Furthermore, the \textbf{Specificity} is the proportion of true negative points to negative elements, as calculated using the equation:
	
	\begin{equation}\label{Specificity}
		\small
		\begin{aligned}
			Specificity =&  \frac{TN}{(TN+FP)}
		\end{aligned}
	\end{equation}
	
	The false positive rate (FPR), represents the ROC curve "X-axis", as calculated using the equation\cite{elhamahmy2010new}:
	
	\begin{equation}\label{FPR}
		\small
		\begin{aligned}
			FPR = 1 - Specificity =&  \frac{FP}{(TN+FP)}
		\end{aligned}
	\end{equation}
	
	In addition, \textbf{Recall} or true positive rate is the ratio of correctly predicted positive observations to all observations in the actual class. True positive rate represents the ROC curve's "Y-axis"
	
	The equation for Recall is shown below.
	
	\begin{equation}\label{Recall}
		\small
		\begin{aligned}
			Recall =&  \frac{TP}{(TP+FN)}
		\end{aligned}
	\end{equation}
	
	The false negative rate (FNR) is calculated using the equation\cite{elhamahmy2010new}:
	
	\begin{equation}\label{FNR}
		\small
		\begin{aligned}
			FNR = 1 - Recall =&  \frac{FN}{(TP+FN)}
		\end{aligned}
	\end{equation}
	
	
		
		
		Finally, the area under the curve (AUC) - receiver operating characteristics (ROC)  (AUC - ROC) plot is another indicator that is used to evaluate the performance of classification models. The Receiver Operating Characteristics (ROC) of a classifier shows its performance as a trade off between False Positive Rate and True Positive Rate. 
		
		In this study, we performed a total of 160 different simulations for each scenario. We performed simulations with changes to the ranges of membership functions of combination ETX and Retransmissions, combination of PDPT \& Retransmissions, combinations of PDR and Retransmissions, combinations of ETX and PDR, combinations of ETX \& PDPT and finally, we performed corrections into the fuzzy rules of the model. In the following paragraphs we demonstrate the positions that the Sink is placed, the attacks that are executed in the selected place and the resulting attach identification accuracy rate. 
		
		\subsection{Results of the Different Approach}
		
		According to the performance measure of TP, TN, FP and FN we calculated the Accuracy rate,  the Precision rate, the Specificity, the FPR rate, Recall rate and the FNR rate where shown in the table \ref{Results_of_the_Different_approaches}.

		\begin{table*}[h!]
			\centering
			\caption{Results of the Different Approaches}\label{Results_of_the_Different_approaches}
			\begin{adjustbox}{width=\linewidth,center}
				\begin{tabular}{|l|ccc|ccc|ccc|ccc|ccc|ccc|}
					\hline
					\multicolumn{1}{|c|}{\multirow{2}{*}{\textbf{Approach}}}                    & \multicolumn{3}{c|}{\textbf{Accuracy rate}}                                    & \multicolumn{3}{c|}{\textbf{Precision rate}}                                   & \multicolumn{3}{c|}{\textbf{Specificity}}                                      & \multicolumn{3}{c|}{\textbf{FPR rate}}                                         & \multicolumn{3}{c|}{\textbf{Recall}}                                           & \multicolumn{3}{c|}{\textbf{FNR}}                                              \\ \cline{2-19} 
					\multicolumn{1}{|c|}{}                                                      & \multicolumn{1}{c|}{Middle}  & \multicolumn{1}{c|}{Top middle} & Top-left edge & \multicolumn{1}{c|}{Middle}  & \multicolumn{1}{c|}{Top middle} & Top-left edge & \multicolumn{1}{c|}{Middle}  & \multicolumn{1}{c|}{Top middle} & Top-left edge & \multicolumn{1}{c|}{Middle}  & \multicolumn{1}{c|}{Top middle} & Top-left edge & \multicolumn{1}{c|}{Middle}  & \multicolumn{1}{c|}{Top middle} & Top-left edge & \multicolumn{1}{c|}{Middle}  & \multicolumn{1}{c|}{Top middle} & Top-left edge \\ \hline
					\textbf{\begin{tabular}[c]{@{}l@{}}ETX and\\ Retransmissions\end{tabular}}  & \multicolumn{1}{c|}{95\%}    & \multicolumn{1}{c|}{93.51\%}    & 93.56\%       & \multicolumn{1}{c|}{87.93\%} & \multicolumn{1}{c|}{78.47\%}    & 76.06\%       & \multicolumn{1}{c|}{96.56\%} & \multicolumn{1}{c|}{94.29\%}    & 93.37\%       & \multicolumn{1}{c|}{3.43\%}  & \multicolumn{1}{c|}{5.70\%}     & 6.62\%        & \multicolumn{1}{c|}{89.40\%} & \multicolumn{1}{c|}{90.13\%}    & 90.13\%       & \multicolumn{1}{c|}{10.59\%} & \multicolumn{1}{c|}{9.86\%}     & 5.57\%        \\ \hline
					\textbf{\begin{tabular}[c]{@{}l@{}}PDPT and\\ Retransmissions\end{tabular}} & \multicolumn{1}{c|}{95.10\%} & \multicolumn{1}{c|}{93.64\%}    & 92.68\%       & \multicolumn{1}{c|}{88.08\%} & \multicolumn{1}{c|}{78.81\%}    & 73.14\%       & \multicolumn{1}{c|}{96.6\%}  & \multicolumn{1}{c|}{94.39\%}    & 92.26\%       & \multicolumn{1}{c|}{3.4\%}   & \multicolumn{1}{c|}{5.60\%}     & 7.73\%        & \multicolumn{1}{c|}{89.76\%} & \multicolumn{1}{c|}{90.41\%}    & 94.57\%       & \multicolumn{1}{c|}{10.23\%} & \multicolumn{1}{c|}{9.58\%}     & 5.42\%        \\ \hline
					\textbf{\begin{tabular}[c]{@{}l@{}}PDR and \\ Retransmissions\end{tabular}} & \multicolumn{1}{c|}{93.80\%} & \multicolumn{1}{c|}{91.69\%}    & 91.61\%       & \multicolumn{1}{c|}{78.99\%} & \multicolumn{1}{c|}{71.17\%}    & 69.09\%       & \multicolumn{1}{c|}{92.73\%} & \multicolumn{1}{c|}{91.25\%}    & 90.25\%       & \multicolumn{1}{c|}{7.26\%}  & \multicolumn{1}{c|}{8.75\%}     & 9.74\%        & \multicolumn{1}{c|}{97.61\%} & \multicolumn{1}{c|}{93.61\%}    & 97.71\%       & \multicolumn{1}{c|}{2.38\%}  & \multicolumn{1}{c|}{6.38\%}     & 2.28\%        \\ \hline
					\textbf{\begin{tabular}[c]{@{}l@{}}ETX and \\ PDR\end{tabular}}             & \multicolumn{1}{c|}{80.36\%} & \multicolumn{1}{c|}{85.46\%}    & 84.66\%       & \multicolumn{1}{c|}{60.09\%} & \multicolumn{1}{c|}{70.76\%}    & 60.53\%       & \multicolumn{1}{c|}{94.33\%} & \multicolumn{1}{c|}{96.34\%}    & 93.37\%       & \multicolumn{1}{c|}{5.66\%}  & \multicolumn{1}{c|}{3.65\%}     & 6.62\%        & \multicolumn{1}{c|}{30.47\%} & \multicolumn{1}{c|}{38.33\%}    & 45.57\%       & \multicolumn{1}{c|}{69.52\%} & \multicolumn{1}{c|}{61.66\%}    & 54.42\%       \\ \hline
					\textbf{\begin{tabular}[c]{@{}l@{}}ETX and \\ PDPT\end{tabular}}            & \multicolumn{1}{c|}{76.27\%} & \multicolumn{1}{c|}{86.51\%}    & 84.14\%       & \multicolumn{1}{c|}{45.63\%} & \multicolumn{1}{c|}{67.78\%}    & 55.63\%       & \multicolumn{1}{c|}{85.26\%} & \multicolumn{1}{c|}{94.13\%}    & 88.59\%       & \multicolumn{1}{c|}{14.73\%} & \multicolumn{1}{c|}{5.8\%}      & 11.4\%        & \multicolumn{1}{c|}{44.16\%} & \multicolumn{1}{c|}{53.47\%}    & 64.14\%       & \multicolumn{1}{c|}{55.83\%} & \multicolumn{1}{c|}{46.52\%}    & 35.85\%       \\ \hline
				\end{tabular}
			\end{adjustbox}
		\end{table*}

		\subsection{Comparison of Results Between each Investigated Approach}
		\label{comparisonofresults}
		
\begin{figure} [h!]
			\centering
			\includegraphics[width=0.99\linewidth]{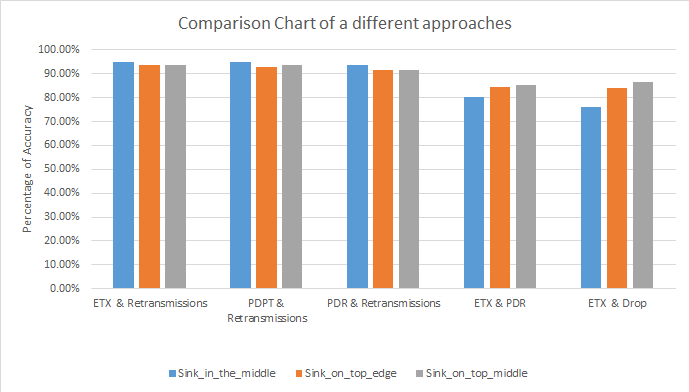}
			\caption{Comparison Chart of a different approaches}
			\label{Comparison_Chart_of_a_different_approaches}
		\end{figure}

		In order to find the optimum combination of the input metrics, we evaluate each solution based on the accuracy rate, the Precision rate, the Specificity, the FPR rate, Recall rate and, the FNR rate. 
		Based on the results, the three combinations 1) ETX  \& Retransmissions, 2) PDPT \& Retransmissions and 3) PDR \& Retransmissions are the optimum inputs for the fuzzy controller. Figure \ref{Comparison_Chart_of_a_different_approaches} shows the Detection Accuracy of the different scenarios of each combination of input metrics. 
		The combinations of ETX \& Retransmissions and PDPT \& Retransmissions achieve the best accuracy of 95\%. Furthermore, we generate the ROC curve of our approach.

\begin{figure} [h!]
			\centering
			\includegraphics[width=1\linewidth]{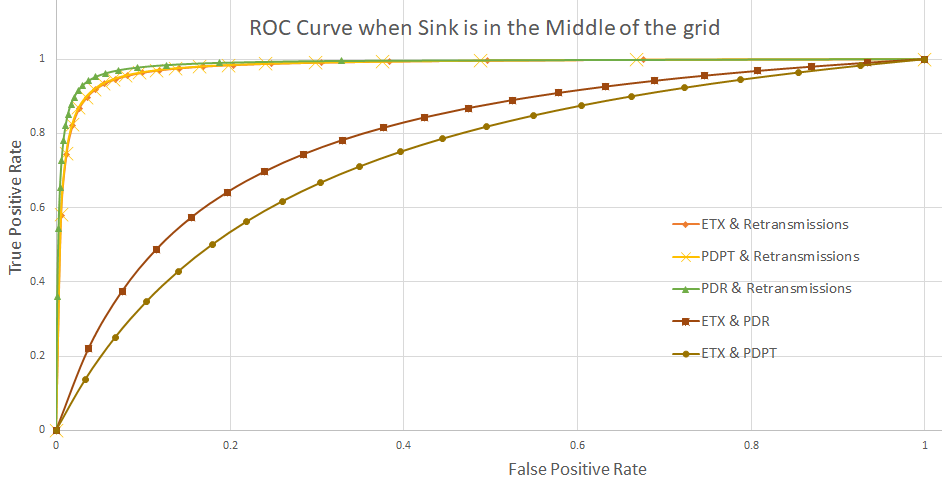}
			\caption{ROC Curve when Sink is in the Middle of the grid}
			\label{ROC_Curve_when_Sink_is_in_the_Middle_of_the_grid}
		\end{figure}

		\begin{table*}[h!]
			\centering
			\tiny
			\caption{The best operating point}\label{best_operating}
			\begin{tabular}{|l|ccc|ccc|}
				
				\hline
				\multicolumn{1}{|c|}{\multirow{2}{*}{\textbf{Approach}}}                    & \multicolumn{3}{c|}{\textbf{Detection Rate}}                                  & \multicolumn{3}{c|}{\textbf{False Alarm Rate}}                                \\ \cline{2-7} 
				\multicolumn{1}{|c|}{}                                                      & \multicolumn{1}{c|}{Middle} & \multicolumn{1}{c|}{Top middle} & Top-left edge & \multicolumn{1}{c|}{Middle} & \multicolumn{1}{c|}{Top middle} & Top-left edge \\ \hline
				\textbf{\begin{tabular}[c]{@{}l@{}}ETX and\\ Retransmissions\end{tabular}}  & \multicolumn{1}{c|}{93\%}   & \multicolumn{1}{c|}{92.8\%}     & 93.4\%        & \multicolumn{1}{c|}{5\%}    & \multicolumn{1}{c|}{8\%}        & 5.6\%         \\ \hline
				\textbf{\begin{tabular}[c]{@{}l@{}}PDPT and\\ Retransmissions\end{tabular}} & \multicolumn{1}{c|}{93\%}   & \multicolumn{1}{c|}{92.95\%}    & 90.9\%        & \multicolumn{1}{c|}{5\%}    & \multicolumn{1}{c|}{7\%}        & 7.9\%         \\ \hline
				\textbf{\begin{tabular}[c]{@{}l@{}}PDR and \\ Retransmissions\end{tabular}} & \multicolumn{1}{c|}{95\%}   & \multicolumn{1}{c|}{92.89\%}    & 94.9\%        & \multicolumn{1}{c|}{4\%}    & \multicolumn{1}{c|}{7.8\%}      & 4.5\%         \\ \hline
				\textbf{\begin{tabular}[c]{@{}l@{}}ETX and \\ PDR\end{tabular}}             & \multicolumn{1}{c|}{74\%}   & \multicolumn{1}{c|}{81.80\%}    & 74.3\%        & \multicolumn{1}{c|}{2.8\%}  & \multicolumn{1}{c|}{21\%}       & 25.6\%        \\ \hline
				\textbf{\begin{tabular}[c]{@{}l@{}}ETX and \\ PDPT\end{tabular}}            & \multicolumn{1}{c|}{66\%}   & \multicolumn{1}{c|}{79.6\&}     & 75\%          & \multicolumn{1}{c|}{3\%}    & \multicolumn{1}{c|}{17\%}       & 26\%          \\ \hline
			\end{tabular}
		\end{table*}
		Figure \ref{ROC_Curve_when_Sink_is_in_the_Middle_of_the_grid} shows the ROC Curve when the sink is placed in the middle of the grid. 
		In figure \ref{ROC_Curve_when_Sink_is_in_the_Middle_of_the_grid}, the combination of ETX \& Retransmissions, PDPT \& Retransmissions and PDR \& Retransmissions are better than the combination of ETX \& PDR and ETX \&  PDPT, because at all cut-offs the true positive rate is higher and the false positive rate is lower. The area under the curve for the combination of ETX \& Retransmissions, PDPT \& Retransmissions and PDR \& Retransmissions is larger than the area under the curve for the combinations ETX \& PDR and ETX \&  PDPT.

		
		
\begin{figure} [h!]
			\centering
			\includegraphics[width=1\linewidth]{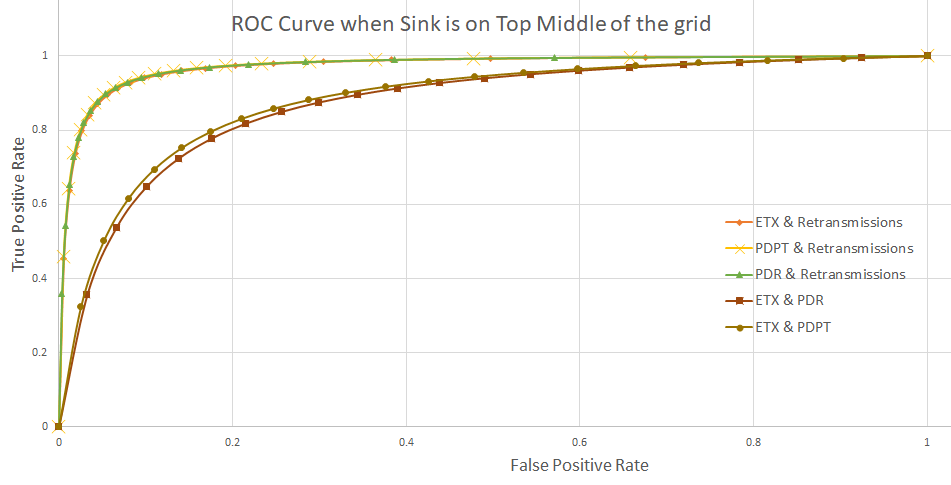}
			\caption{ROC Curve when Sink is on Top Middle of the grid}
			\label{ROC_Curve_when_Sink_is_on_Top_Middle_of_the_grid}
		\end{figure}

		Figure \ref{ROC_Curve_when_Sink_is_on_Top_Middle_of_the_grid} shows the ROC Curve when Sink is placed in the top middle of the grid. Also, in figure \ref{ROC_Curve_when_Sink_is_on_Top_Middle_of_the_grid} the combination of ETX \& Retransmissions, PDPT \& Retransmissions and PDR \& Retransmissions are the best, for the same reasons as in the previous topology.

		
		
\begin{figure} [h!]
			\centering
			\includegraphics[width=1\linewidth]{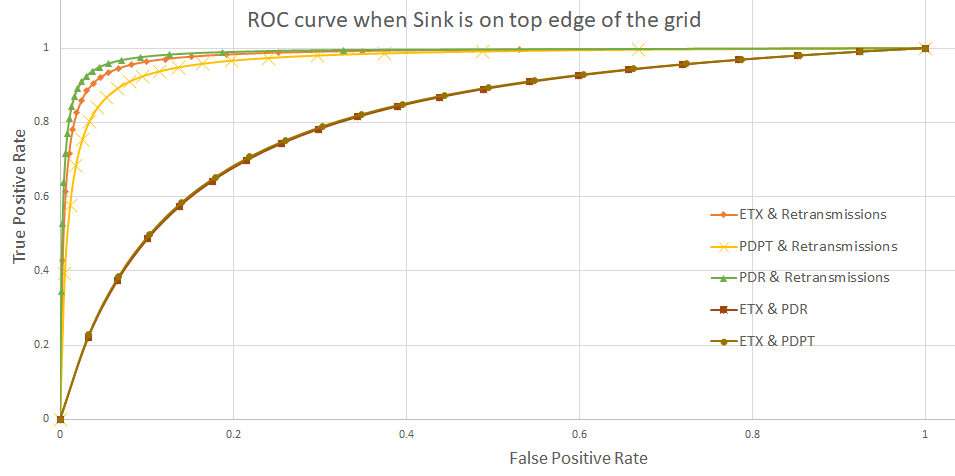}
			\caption{ROC Curve when Sink is on Top Edge of the grid}
			\label{ROC_Curve_when_Sink_is_on_Edge_Middle_of_the_grid}
		\end{figure}

		Figure \ref{ROC_Curve_when_Sink_is_on_Edge_Middle_of_the_grid} shows the ROC Curve when Sink is placed in the Top Edge of the grid. Additionally, in figure \ref{ROC_Curve_when_Sink_is_on_Edge_Middle_of_the_grid} the combination of ETX \& Retransmissions, PDPT \& Retransmissions and PDR \& Retransmissions are still the best.

		

		In conclusion, we summarize the best operating point for each approach in table \ref{best_operating}. According to the table \ref{best_operating} the best approaches are the combination of ETX and Retransmissions where perform locally detection and the combination of PDR and Retransmissions where perform centrally detection.


		\section{Conclusions}
		\label{conclussions_and_future_work} 
		
		Overall, results show that the Retransmissions metric, locally collected at the nodes, is the most suitable value as an input for the fuzzy controller. From the results, we observe that the combination of ETX \& Retransmissions and PDPT  \&  Retransmissions have the best accuracy achieve at 95\%. These set of parameters are having better performance because these parameters are heavily affected by the jamming attacks. Note that both approaches are based on distributed data and decision can be executed at node. Additionally, ROC curves show that the combination of ETX \& Retransmissions, PDPT \& Retransmssions  and  PDR  \&  Retransmissions  are more useful from the combinations ETX \& PDR and ETX \& PDPT. Based on the results of the ongoing work, we have identified several open issues that will be studied as part of this research for future work. 
		
		

		\balance
		
		\bibliography{conference_ieee}
		\bibliographystyle{ieeetr}

	\end{document}